\PassOptionsToPackage{dvipsnames}{xcolor}
\documentclass[fleqn,10pt]{wlscirep}
\usepackage[utf8]{inputenc}
\usepackage[T1]{fontenc}
\usepackage{textgreek}
\usepackage{textcomp}
\DeclareUnicodeCharacter{0413}{\textGamma}
\usepackage{bm}

\usepackage{amsmath,amssymb,amsfonts}
\usepackage{bm}
\usepackage{color, soul}
\usepackage{graphicx}
\usepackage{float}
\usepackage{wrapfig}
\usepackage{newunicodechar}
\newunicodechar{≈}{\approx}
\usepackage{verbatim}
\usepackage[english]{babel}
\usepackage[dvipsnames]{xcolor}

\title{Gas-induced reversible spectral tuning of bound states in the continuum in perovskite metasurfaces}

\author[1,*]{Nikita~Glebov}
\author[2,3,*]{Yuliya~Kenzhebayeva}
\author[3]{Pavel~Pankin}
\author[1]{Mikhail~Masharin}
\author[3]{Martin~Rojas-Bustamante}
\author[2]{Xiuzhen~Liu}
\author[2]{Shixiong~Li}
\author[2,3]{Ruslan~Azizov}
\author[4]{Dmitrii~Maksimov}
\author[1]{Hatice~Altug}
\author[2,3,**]{Andrey~Bogdanov}
\author[2,3,**]{Sergey~Makarov}

\affil[1]{{Institute of Bioengineering}, {\'Ecole Polytechnique F\'ed\'erale de Lausanne (EPFL)}, {{Lausanne}, {1015}, {Switzerland}}}
\affil[2]{{Qingdao Innovation and Development Center}, {Harbin Engineering University}, {{Qingdao}, {266000}, {China}}}
\affil[3]{{School of Physics and Engineering}, {ITMO University}, {St. Petersburg}, {197101}, {Russia}}
\affil[4]{Kirensky Institute of Physics, Federal Research Center KSC SB
RAS, 660036, Krasnoyarsk, Russia}

\affil[*]{These authors contributed equally to this work}
\affil[**]{e-mail: a.bogdanov@metalab.ifmo.ru, s.makarov@metalab.ifmo.ru}

\begin{document}

\flushbottom

\begin{abstract}

   Bound states in the continuum (BICs) provide a powerful route to engineering optical resonances in compact photonic structures, enabling enhanced light--matter interaction for sensing, nonlinear optics, optical modulation, and lasing. The next step towards practical BIC-based meta-optics is dynamic and reversible spectral control after fabrication. However, optical and electro-optical tuning typically produces limited refractive-index changes, thermo-optical approaches require continuous external actuation, and phase-change materials often provide discrete switching rather than continuous spectral reconfiguration. Here we demonstrate giant, fully reversible, and continuous spectral tuning of BIC resonances in halide-perovskite metasurfaces. Using nanoimprint lithography, we fabricate MAPbBr$_3$ metasurfaces, that support BIC-derived resonances in the visible spectral range. Exposure to HCl vapor initiates an anion-exchange reaction in which Cl$^-$ ions diffuse into the perovskite crystalline lattice and partially replace Br$^-$ ions, producing a monotonic increase of the electronic bandgap and blueshift of the excitonic resonance on a minute timescale. This leads to a decrease in the refractive index and tunes the BIC resonance by more than 40~nm in the visible range. Subsequent exposure to HBr vapor reverses the exchange process and restores the resonance to its initial spectral position. As a functional demonstration, we show continuous reversible spectral tuning of BIC-mediated lasing from the same metasurface. Our results establish gas-phase anion exchange as a non-destructive mechanism for continuous, reversible, and large-range post-fabrication control of perovskite metasurfaces, opening opportunities for chemically programmable lasers, adaptive meta-optics, and active nanophotonic devices.
\end{abstract}

\maketitle
\thispagestyle{empty}

\section*{Introduction}

Bound states in the continuum (BICs) have become a fundamental concept in resonant photonics, enabling optical modes with strongly suppressed radiation losses in open systems~\cite{hsu2016bound, PhysRevLett.100.183902, PhysRevLett.107.183901}. In dielectric metasurfaces, BICs generally arise from symmetry protection or destructive interference between radiation channels and remain decoupled from the far field despite lying within the radiation continuum. When weakly perturbed, these states transform into quasi-BIC resonances that retain high quality factors (Q-factors) while becoming experimentally accessible~\cite{PhysRevLett.121.193903, KOSHELEV2019836, https://doi.org/10.1002/lpor.202200564}. This combination of strong field localization, planar geometry, and far-field addressability has made BIC-metasurfaces a versatile platform for enhanced light--matter interaction with applications in sensing, nonlinear optics, optical modulation, and lasing~\cite{gu2023reconfigurable,tonkaev2022multifunctional,khaliq2026recent}.

Such applications of BIC-metasurfaces as sensing~\cite{tittl2018imaging, tan2021active}, adaptive optics~\cite{cotrufo2024reconfigurable}, optical telecommunication~\cite{murai2022engineering}, information encryption~\cite{xu2025dynamic}, and neuromorphic computing~\cite{fu2026photonic} can employ dynamic spectral tuning, unlocking new advanced device functionalities. However, BIC resonance frequencies are primarily determined by metasurfaces' geometrical sizes, symmetry, and constituent materials, making them largely immutable after fabrication. In many cases, BIC-based meta-optics therefore requires tuning mechanisms that are reversible, broadband, and continuously adjustable while preserving the ultrahigh-$Q$ nature of the resonance. Achieving these requirements remains a major challenge in reconfigurable nanophotonics. 

Existing approaches to tunable BIC-metasurfaces address this challenge only partially. All-optical modulation based on free-carrier generation~\cite{berte2025all,brikh2026mid} or polariton nonlinearities~\cite{ardizzone2022polariton,masharin2024giant} can provide ultrafast response, but usually relies on relatively small refractive-index changes suitable for modulator development only. Thermo-optical tuning can yield larger changes, yet typically requires continuous external actuation and introduces constraints associated with heat dissipation, thermal stability, and device integration~\cite{brikh2026mid,meiler2025spectral,liu2026optical}. Phase-change materials provide substantial optical contrast and enable reconfigurable metasurfaces~\cite{cotrufo2024reconfigurable}, but they often rely on discrete or hysteretic switching rather than continuous spectral reconfiguration. These limitations motivate alternative routes in which the resonant material itself can be modified volumetrically and reversibly.

Gas-phase chemical reconfiguration is attractive for tunable metasurfaces because it is contactless, selective, and compatible with post-fabrication processing. It has been used to modify the optical environment of metasurfaces, for example, through liquid-crystal reorientation~\cite{kim2021holographic}, and to tune plasmonic components through reversible gas absorption, as in palladium nanoantennas for hydrogen sensing~\cite{liu2011nanoantenna}. However, extending this concept to all-dielectric high-Q metasurfaces is challenging, because conventional dielectrics rarely exhibit large, reversible, and volumetric refractive-index changes under mild gas exposure.

\begin{figure}[ht]
\centering
\includegraphics[width=0.99\linewidth]{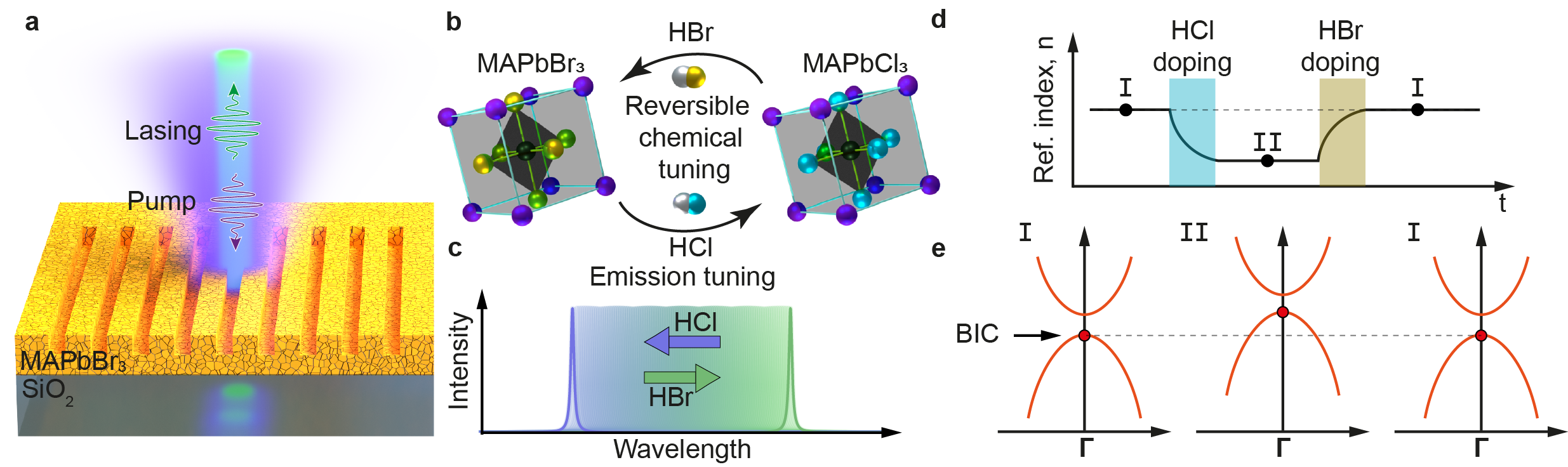}
\caption{
\textbf{Concept of chemically reconfigurable perovskite metasurface lasers.}
\textbf{a,}~Schematic illustration of a nanoimprinted MAPbBr$_3$ grating metasurface on a SiO$_2$ substrate, with lasing under optical pumping.
\textbf{b,}~Gas-phase anion exchange in the perovskite lattice: exposure to HCl vapour replaces Br$^-$ with Cl$^-$ in the PbX$_6$ octahedra; exposure to HBr vapour reverses the exchange.
\textbf{c,}~Reversible spectral tuning of the lasing emission.
\textbf{d,}~Non-volatile tuning of the perovskite refractive index $n$ by anion exchange: state I, pristine MAPbBr$_3$; state II, after HCl exposure. 
\textbf{e,}~Schematic polariton dispersions of the metasurface in states I and II from \textbf{d}. The lower $n$ in state II blue-shifts the modes. Red dots mark the symmetry-protected BIC at $k_x = 0$.
}
\label{fig1}
\end{figure}

Here we demonstrate giant, fully reversible, and continuous spectral reconfiguration of BIC resonances in MAPbBr$_3$ metasurfaces through gas-phase anion exchange (\autoref{fig1}a-c). The BIC-derived resonances undergo a wavelength shift exceeding 40~nm across the visible spectrum upon exposure to HCl vapor, representing a post-fabrication tuning range inaccessible to conventional dielectric metasurfaces. This reconfigurability originates from volumetric halide exchange, where  Cl$^-$ ions diffuse throughout the perovskite crystalline lattice and partially replace Br$^-$ ions, inducing simultaneous modification of the electronic bandgap, excitonic transition, and reducing the refractive index~\cite{nedelcu2015fast,doi:10.1021/acsnano.2c11013}. Reversal with HBr vapor restores the original halide composition and returns the optical resonance to its initial spectral position, demonstrating non-destructive and repeatable control (\autoref{fig1}d,e). Beyond the resonance tuning, we exploit this chemically reconfigurable platform to achieve reversible spectral tuning of polariton lasing from the same metasurface. These results establish gas-phase anion exchange as a new paradigm for large-range, continuous, and reversible post-fabrication control of active perovskite metasurfaces, enabling chemically programmable lasers, adaptive meta-optics, and dynamically reconfigurable nanophotonic systems.
\section*{Results}

\subsection*{Perovskite metasurface supporting bound states in the continuum}

\begin{figure}[ht]
\centering
\includegraphics[width=0.8\linewidth]{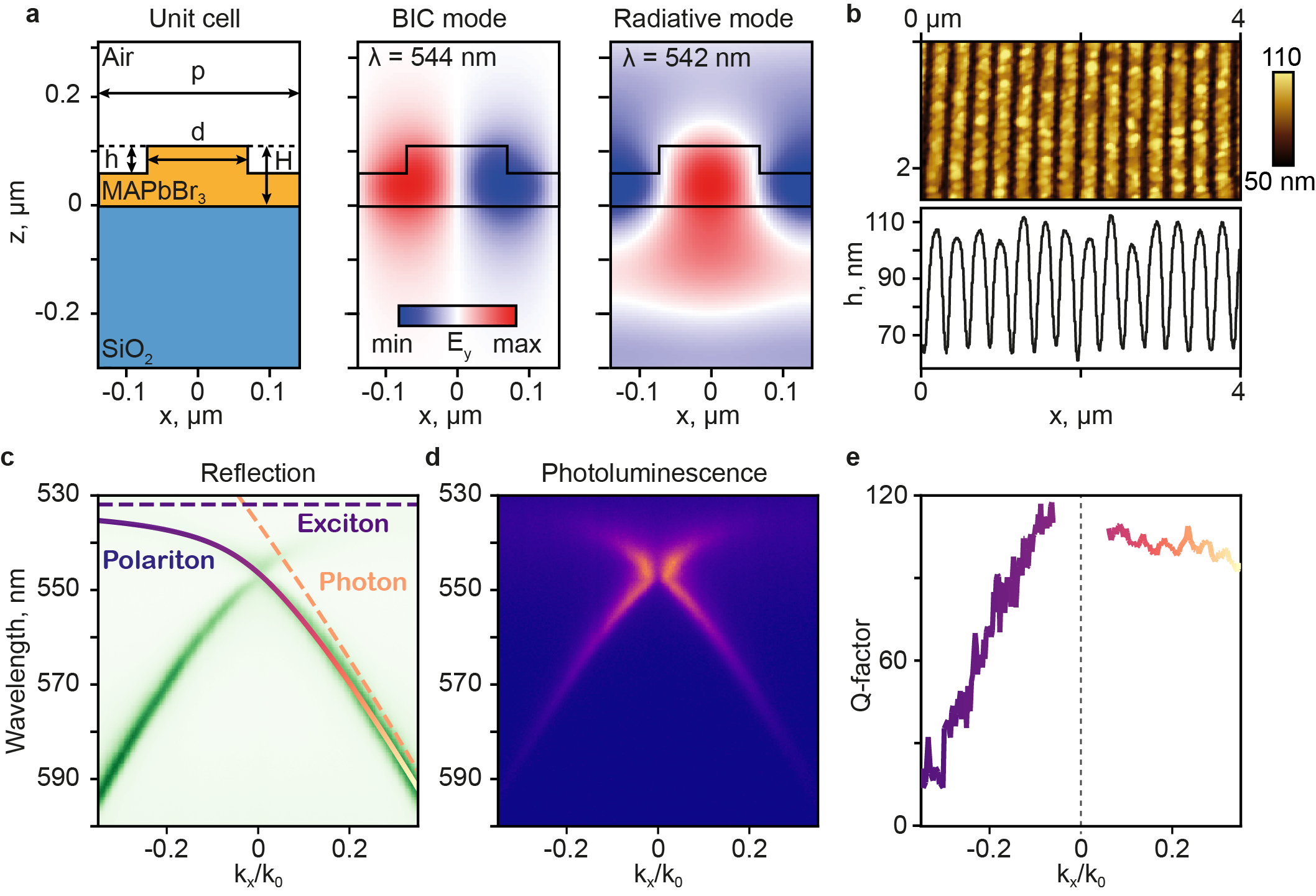}
\caption{
\textbf{Reconfigurable perovskite metasurfaces supporting bound states in the continuum.}
\textbf{a,} Schematic illustration of the perovskite metasurface unit cell. Simulated electric-field distributions of the BIC mode and the radiative mode.
\textbf{b,} Atomic force microscopy (AFM) image of the fabricated perovskite metasurface and the corresponding extracted height profile.
\textbf{c,} Angle-resolved reflection spectra of the perovskite metasurface. Color-coded traces indicate the fitted polariton mode energies. Dashed lines show fitted exciton, photon, and polariton dispersions.
\textbf{d,} Angle-resolved photoluminescence (PL) spectra of the perovskite metasurface.
\textbf{e,} Experimental quality factor ($Q$) of the polariton mode as a function of in-plane wavevector.
}
\label{fig2}
\end{figure}

 We designed dielectric metasurfaces supporting symmetry-protected BICs to realize a high-$Q$ exciton--polariton platform with controllable radiative losses. In an inversion-symmetric periodic grating, the BIC emerges at the $\Gamma$-point due to symmetry mismatch between the confined mode and radiative channels, resulting in vanishing far-field coupling in the ideal structure. We optimized the grating period, modulation depth, and film thickness to set the spectral position and radiative leakage of the BIC resonance, and to overlap it with the gain spectrum of MAPbBr$_3$ for lasing.  We employed a one-dimensional subwavelength grating with a fill factor of 0.5, chosen for fabrication symmetry and robustness (\autoref{fig2}), ensuring reproducible feature definition while supporting spectrally sharp quasi-BIC resonances in the perovskite metasurface. The optimized structure supports both a BIC-type mode and a radiative mode at the $\Gamma$-point. The eigenvalue problem and the scattering problem were solved using the finite difference frequency domain (FDFD) method (see Methods). \autoref{fig2}a shows the field profiles of two eigenmodes at the $\Gamma$-point near 550 nm. The dipole distribution of the electric field of the first eigenmode is confined within the perovskite grating, which corresponds to a symmetry-protected BIC~\cite{PhysRevLett.121.193903}. In contrast, the electric field distribution of the second mode exhibits a radiation pattern in both the substrate and the superstrate, which corresponds to an ordinary resonant state or quasi-normal mode~\cite{sauvan2022normalization}.

To realize a reconfigurable perovskite metasurface, we combined a solution-based method with nanoimprint lithography~\cite{ doi:10.1021/acsnano.4c18847, makarov2017multifold} (see Methods and Supplementary Information (SI) Fig. S1). The resulting MAPbBr$_3$ thin films exhibited a thickness of approximately 110 nm. Nanoimprint master molds were fabricated by electron-beam lithography on fused silica substrates followed by lift-off process of Cr/Au, yielding structures with a period $p = 280$~nm, a line width $d = 140$~nm, and a modulation depth $h = 50$~nm (see Methods). We imprinted the perovskite film in its intermediate phase under a stationary pressure at room temperature, and then annealed it as described in the Methods. The fabrication process yields a 70$\times$50~$\mu$m size perovskite metasurface with faithful reproduction of the mold geometry, as confirmed by atomic force microscopy (AFM) (\autoref{fig2}b). The extracted height profile shows a modulation depth of $h \approx 50$~nm. Importantly, the perovskite material remains structurally intact throughout the nanoimprint lithography process, thereby preserving its optical properties.

Angle-resolved reflection and photoluminescence (PL) measurements in TE polarization were performed using a back-focal-plane (BFP) 4$f$ imaging setup~\cite{10.1063/5.0142570, doi:10.1126/sciadv.abj6627} (see Methods and SI Fig. S2) to probe the optical response of the perovskite metasurface. Angle-resolved reflection spectra (\autoref{fig2}c) reveals the characteristic anticrossing behavior of the modes near the exciton resonance, confirming the strong light--matter coupling regime at room temperature~\cite{deng2010exciton}. The extracted dispersion (\autoref{fig2}c) was fitted using a coupled two-oscillator model to quantify the exciton--polariton parameters (see SI, note 3), yielding a Rabi splitting of 116~meV and an exciton resonance energy corresponding to MAPbBr$_3$ at 532~nm~\cite{doi:10.1021/acsphotonics.2c01773}. The resulting polariton dispersion is shown as the colour-coded fitted curve in \autoref{fig2}c. Angle-resolved PL measurements (\autoref{fig2}d) reveal a pronounced suppression of emission at $k_x/k_0=0$, consistent with the formation of a symmetry-protected BIC\cite{wu2024exciton}. The experimentally extracted quality factors increase towards the $\Gamma$-point for both modes. For the lower polariton mode, this enhancement originates from the vanishing radiative coupling of the photonic BIC component. For the upper polariton mode, the increased quality factor arises from the reduced excitonic contribution near the $\Gamma$-point, which suppresses exciton-related losses. Since excitonic dissipation dominates over photonic losses in this system, the reduction of the excitonic fraction results in reduced linewidth broadening near the $\Gamma$-point (\autoref{fig2}e).

\subsection*{Gas tuning of metasurface}

To achieve giant and continuous spectral tuning of the BIC resonances, we exploit gas-phase anion exchange in the perovskite metasurface through exposure to HCl vapour. This process initiates the incorporation of Cl$^-$ ions into the perovskite lattice, where they progressively substitute Br$^-$ ions. To quantitatively describe the exchange kinetics, we developed a diffusion model that captures the coupled processes of surface anion exchange and subsequent Cl$^-$ transport within the perovskite metasurface. Unlike previous studies performed on single-crystalline nanowires~\cite{doi:10.1021/acsnano.2c11013}, where ionic transport is dominated by bulk diffusion, our metasurface consists of a polycrystalline MAPbBr$_3$ film. In this case, grain boundaries provide additional low-energy diffusion pathways that enhance halide migration~\cite{shao2016grain}. Therefore, we describe Cl$^-$ transport using an effective diffusion coefficient that incorporates contributions from both lattice and grain-boundary diffusion.
\begin{figure}[ht!]
\centering
\includegraphics[width=0.8\linewidth]{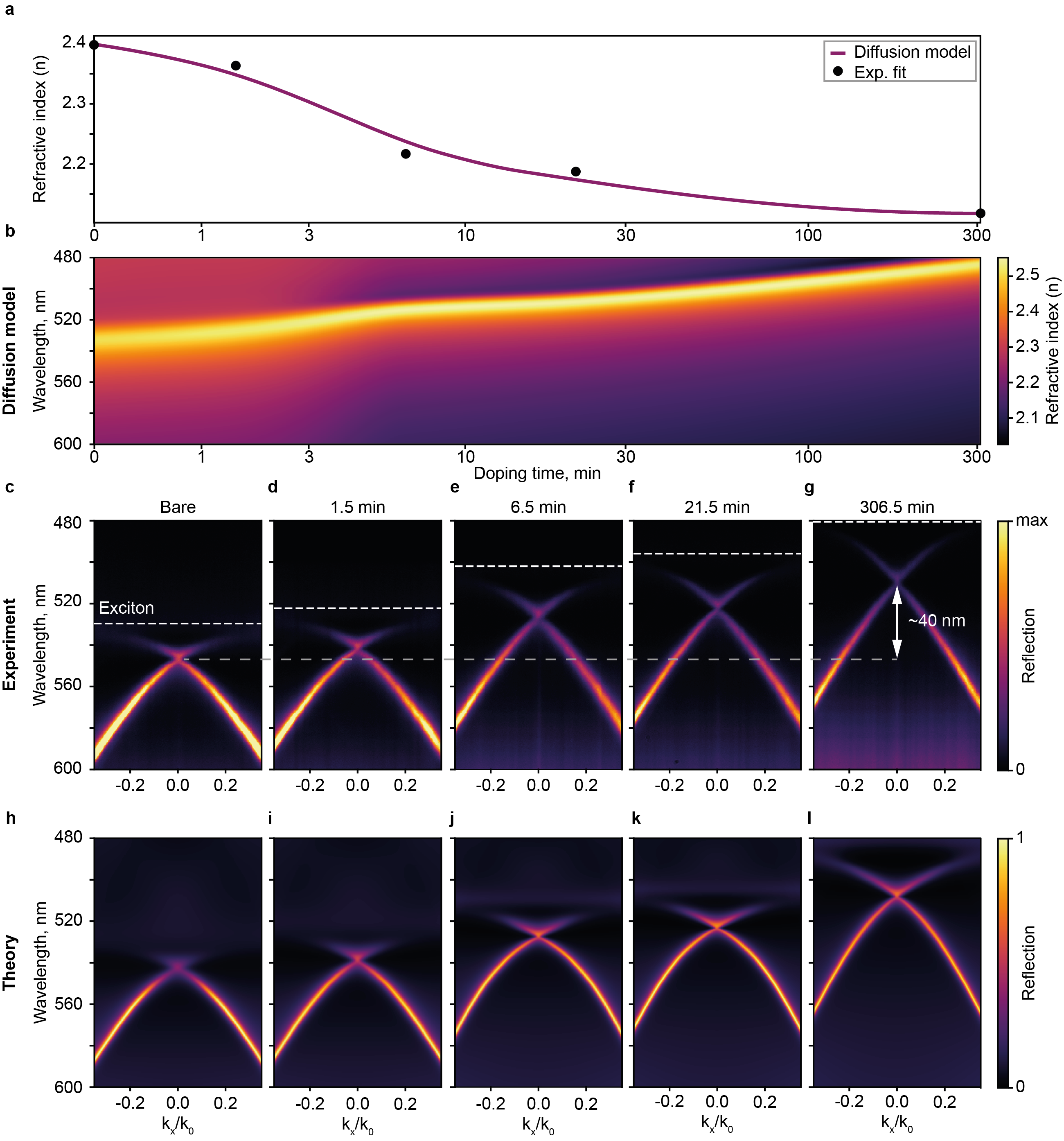}
\caption{
\textbf{Gas tuning of polaritonic BIC modes.}
\textbf{a,} Refractive index at a wavelength of 545 nm as a function of doping time, calculated using the diffusion model (violet curve) and estimated from experimental reflectivity measurements (black dots). 
\textbf{b,}  Spectral and temporal dependence of the refractive index calculated using the diffusion model.
\textbf{c--g,} Experimental angle-resolved reflection spectra of perovskite metasurfaces recorded after gas doping for 0, 1.5, 6.5, 21.5 and 306.5 min, respectively.
\textbf{h--l,} Corresponding simulated angle-resolved reflection spectra. }
\label{fig3}
\end{figure}
The model assumes that HCl molecules reaching the exposed perovskite surfaces undergo an anion-exchange reaction, where Br$^-$ ions are continuously replaced by Cl$^-$ ions while maintaining the corner-sharing PbX$_6$ octahedral framework of the perovskite crystal. The exchange reaction is confined to the interfaces directly exposed to the gas environment, namely the top surface and the sidewalls of the grating ridges. Following incorporation, Cl$^-$ ions diffuse into the perovskite volume driven by concentration gradients established by the surface exchange process. Our approach accounts for both vertical Cl$^-$ penetration from the top surface and lateral diffusion from the exposed ridge sidewalls.  As the exchange proceeds, the composition continuously evolves from MAPbBr$_3$ towards MAPbBr$_{3-x}$Cl$_x$, resulting in an increased bandgap and a corresponding reduction of the refractive index. By fitting the temporal evolution of the optical response, we extract an effective diffusion coefficient of $1.0 \times 10^{-10}$~cm$^2$\,s$^{-1}$. At this diffusion rate, the refractive-index distribution remains nearly homogeneous throughout the metasurface during the tuning process (see SI Fig. S3).

We exposed the fabricated perovskite metasurface to HCl vapor inside a sealed Petri dish for controlled durations to investigate the dynamic tuning of the metasurface resonances (see Methods and SI Fig. S1). The HCl vapor atmosphere was generated by evaporating a droplet of hydrochloric acid at 50~$^\circ$C. After establishing the vapor environment, the samples were exposed for times ranging from 0 to 306.5 min. Following each exposure step, angle-resolved reflection and PL spectra were recorded using the same BFP setup (\autoref{fig3}b--f). With increasing exposure time, the exciton resonance, exciton--polariton branches, and, in particular, the quasi-BIC resonance progressively shift towards shorter wavelengths. This continuous blue shift originates from the gradual substitution of Br$^-$ by Cl$^-$ ions within the perovskite lattice, which increases the electronic bandgap and reduces the refractive index of the material (SI {Fig.~S4}). After 306.5 min of exposure, the total spectral blueshift exceeds 40 nm, demonstrating exceptionally large post-fabrication reconfiguration of an all-dielectric metasurface operating in the visible spectral range. Notably, the resonance remains spectrally well-defined throughout the entire tuning process without measurable mode splitting or additional linewidth broadening. This observation indicates that the ion diffusion is sufficiently rapid to maintain a nearly homogeneous refractive-index distribution across the metasurface, thereby preserving both the grating morphology and the high-Q character of the resonance.

To support the experimental observations, we calculated the angle-resolved reflectance spectra (\autoref{fig3}g--k) using the FDFD (see Methods), employing the refractive-index profiles obtained from the diffusion model (see SI Fig. S5). The simulated spectra are in good agreement with the experimental results and reproduce the characteristic avoided crossing of the exciton--polariton branches. At the $\Gamma$-point, one branch collapses into a symmetry-protected BIC, confirming the designed modal behavior of the metasurface. As the doping time increases, the progressive reduction of the perovskite refractive index induces a continuous blue shift of the photonic resonance. Consequently, both exciton-polariton modes shift towards shorter wavelengths, in excellent agreement with the experimental observations. The simultaneous blue shift of the excitonic absorption resonance reduces the material absorption at the resonance wavelength, resulting in lower non-radiative losses and an increase in the material quality factor for both modes. In contrast, the radiative losses of the upper mode increase with increasing Cl content, leading to a reduction of its total quality factor.
A detailed analysis of the individual radiative and material loss contributions is provided in SI {Fig.S6-S7}.
Furthermore, the Zel'dovich perturbation theory presented in (SI, note 7) provides a general framework for predicting the evolution of the complex eigenfrequency under gas-induced permittivity variations.
This approach enables simultaneous evaluation of both the resonance wavelength and its linewidth, offering two complementary observables that can be exploited for BIC tuning and optical sensing.

\subsection*{Reversible gas tuning of lasing}

\begin{figure}[ht!]
\centering
\includegraphics[width=0.8\linewidth]{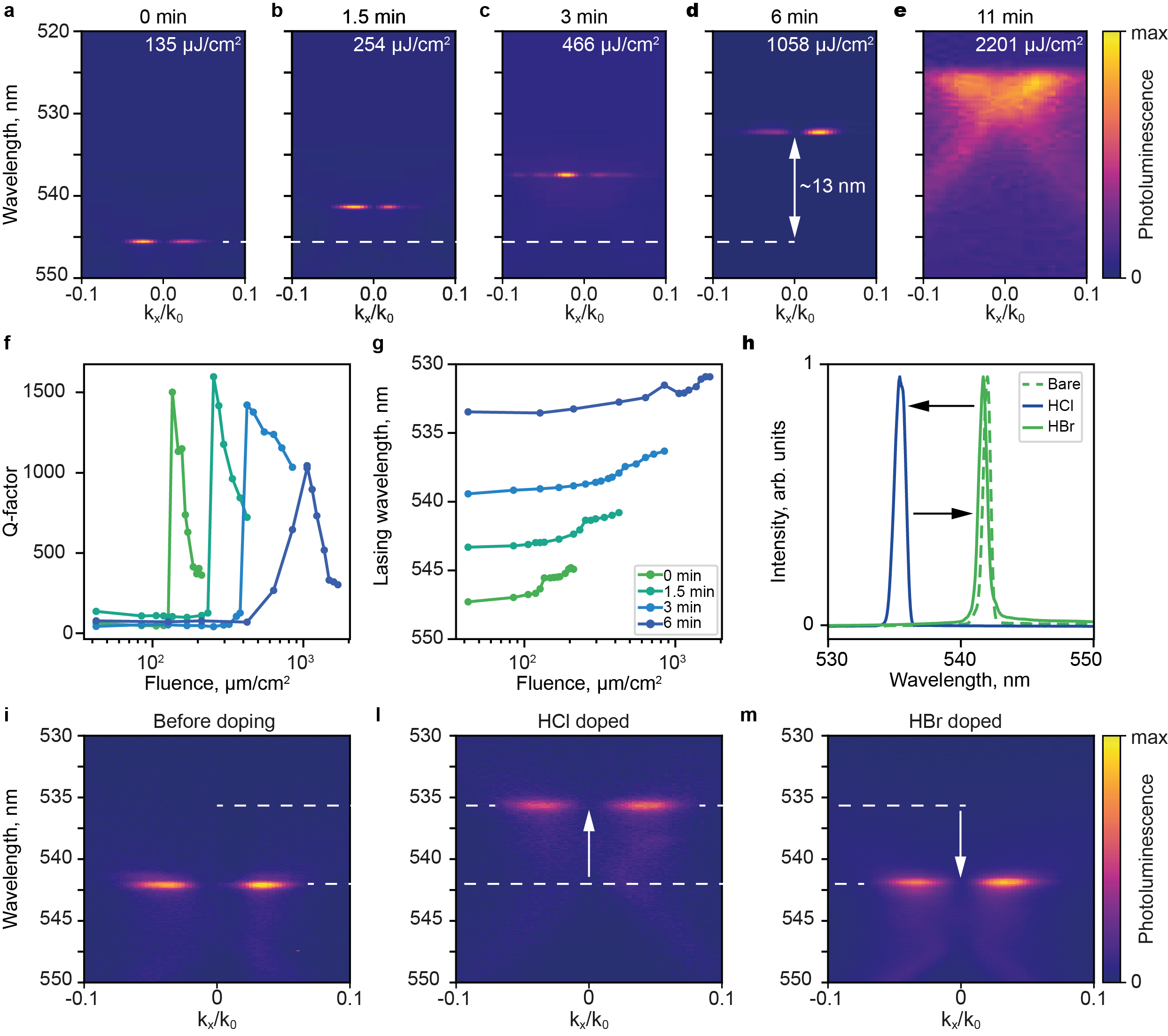}
\caption{\textbf{Reversible tuning of lasing via gas doping.} \textbf{a--e,} Angle-resolved photoluminescence spectra of perovskite metasurfaces recorded after different gas-phase doping durations. Lasing is observed in panels a--d, whereas only photoluminescence remains in e. \textbf{f,} Q-factor of the lasing mode as a function of pump fluence. \textbf{g,} Extracted lasing wavelength as a function of pump fluence. \textbf{h,} Lasing intensity profiles for pristine MAPbBr$_3$, after HCl doping, and following recovery by HBr doping. \textbf{i--m,} Corresponding angle-resolved photoluminescence spectra. }
\label{fig4}
\end{figure}

The combination of optical gain and high-quality resonances makes the perovskite metasurface an attractive platform for polariton-mediated lasing~\cite{masharin2023room}. To investigate the stimulated emission characteristics, we performed angle-resolved photoluminescence spectroscopy as a function of excitation fluence for metasurfaces with different halide compositions, where the light source was a femtosecond laser (see Methods). Pristine MAPbBr$_3$ metasurfaces exhibit a pronounced nonlinear increase in emission intensity accompanied by a rapid linewidth narrowing, indicating the onset of polariton lasing at a pump fluence of 135~$\mu J/cm^2$ (see SI Fig.~S9). With increasing pump fluence, the intensity dip at the $\Gamma$ point, initially present only for the lower mode, extends to both branches. We attribute this to the ultrafast carrier-induced spectral shift of the polariton modes~\cite{schlaus2019lasing, doi:10.1021/acsphotonics.2c01773, fieramosca2019two}, which reduces the detuning between the BIC and the radiative mode and can lead to their hybridization. \autoref{fig4}a--e presents the angle-resolved photoluminescence spectra recorded at pump fluences slightly above the lasing threshold for different HCl exposure times. As the doping time increases, the lasing mode exhibits a continuous blue shift, following the evolution of the exciton--polariton dispersion shown previously. At the same time, the lasing threshold gradually increases. After approximately 11 min of HCl exposure, however, no lasing is observed within the maximum available pump fluence. We attribute the disappearance of lasing to the different evolution of the optical gain spectrum and the cavity resonance during anion exchange. While the cavity resonance is primarily governed by the reduction of the refractive index, the gain spectrum shifts owing to the increasing bandgap of the mixed-halide perovskite. Consequently, the spectral overlap between the $\Gamma$-point mode and the gain maximum is progressively reduced, resulting in increased threshold and eventual suppression of lasing.

To determine the lasing threshold, we extracted the quality factors and emission wavelengths of the polariton modes as a function of excitation fluence by fitting the emission spectra with Lorentzian functions (\autoref{fig4}f,g). Below the threshold, the quality factor remains approximately constant at $Q \approx 100$, indicating spontaneous emission into the quasi-BIC mode. Above a critical pump fluence, the quality factor increases abruptly to $Q \approx 1500$, accompanied by a pronounced linewidth narrowing, which is a hallmark of polariton lasing~\cite{byrnes2014exciton,masharin2023room}. Consistent with the evolution of the gain-cavity detuning discussed above, the lasing threshold increases systematically with HCl exposure time. After only 1.5~min of doping, a modest blue shift of the lasing wavelength from 545.5 to 541.3~nm is observed, corresponding to an energy shift of approximately 18~meV, while the emission linewidth remains nearly unchanged at 0.34~nm. Extending the exposure to 3~min results in a more pronounced shift of $\approx$7.6~nm, accompanied by slight broadening to 0.38~nm.  At the stage of 6 min, the lasing peak has shifted by $\approx$13.4 nm, and the linewidth broadens to 0.52~nm. In addition to the overall blue shift of the lasing wavelength induced by anion exchange, we observed a further pump-dependent blue shift above threshold, corresponding to an energy shift of approximately 3~meV (\autoref{fig4}g). This nonlinear shift is attributed to polariton-polariton interactions and carrier-induced renormalization of the excitonic resonance (polaron effects), consistent with previous reports on perovskite polariton lasers\cite{doi:10.1021/acsphotonics.2c01773}. Importantly, the lasing mode remains stable and does not exhibit mode hopping or degradation under the pump pulses for all doping times, confirming the robustness of the reconfigurable metasurface and the photostability of the MAPbBr$_{3-x}$Cl$_{x}$ gain medium under our excitation conditions.

To demonstrate the reversibility of the proposed tuning mechanism, we exposed the metasurfaces to HBr vapor following the same procedure used for HCl treatment. Under these conditions, Br$^-$ ions diffuse into the perovskite lattice and progressively replace Cl$^-$ ions, driving the composition back towards pristine MAPbBr$_3$. Crucially, HBr exposure fully reverses the anion-exchange process, restoring the lasing peak to its initial spectral position while preserving its linewidth and emission intensity (\autoref{fig4}h). Furthermore, the angle-resolved lasing spectra after HBr treatment are nearly identical to those of the pristine metasurface (\autoref{fig4}i--m), confirming complete reversibility of the spectral tuning and demonstrating the resilience of the perovskite metasurface under repeated gas-phase anion exchange.

\section*{Conclusion}

In conclusion, we have demonstrated a giant and fully reversible gas-induced reconfiguration of a high-Q BIC resonance in an all-dielectric MAPbBr$_3$ metasurface. Through a post-fabrication anion exchange process driven by exposure to HCl and HBr vapors, we achieved a spectral shift of the BIC mode exceeding 40 nm. The complete reversibility of this process has been confirmed by the cyclical switching of the lasing emission peak. Our comprehensive experimental and theoretical analysis reveal that the mechanism is driven by the deep penetration of halide ions into the perovskite lattice, which directly modulates the material's electronic band gap and refractive index. This represents a significant paradigm shift from previous gas-tuning methods, typically limited to plasmonic systems and surface interactions, and establishes a new platform for chemical control in dielectric nanophotonics.

The ability to tune optical BICs through reversible gas-phase anion exchange opens opportunities for reconfigurable resonant photonic systems. In particular, the chemical nature of the tuning mechanism offers an important distinction from approaches that rely on continuous optical or electrical excitation: once the desired compositional state is reached, the corresponding optical response can be maintained without sustained energy input and reconfigured when required. The reversible tuning of lasing directly at the BIC position further highlights the potential of this approach for dynamically reconfigurable light sources and laser displays~\cite{li2026photoinduced}. Extending this concept to other perovskite compositions and gas-phase stimuli could provide access to a broader spectral range and additional functionalities, establishing reversible ion exchange as a versatile route towards reconfigurable perovskite meta-optics.
\newpage
\section*{Methods}
\subsection*{Fabrication}
\subsubsection*{Fabrication of perovskite waveguide metasurfaces.}
The perovskite metasurfaces were fabricated using spin-coating followed by nanoimprint lithography. The fabrication began with preparation of the perovskite precursor solution. Methylammonium bromide (MABr) and lead(II) bromide (PbBr$_2$) were dissolved in a 3:1 mixture of dimethylformamide (DMF) and dimethyl sulfoxide (DMSO) to achieve a 0.5 M solution, with masses $m_{\mathrm{MABr}} = 55.98$ mg and $m_{\mathrm{PbBr_2}} = 183.51$ mg dissolved in 1 mL total solvent. The solution was prepared in a nitrogen-filled glovebox to prevent degradation.

Prior to film deposition, 15 × 15 mm$^2$ SiO$_2$ substrates were sequentially cleaned with detergent solution, acetone, and isopropanol, followed by oxygen plasma treatment for 10 min to improve surface adhesion. The perovskite film was deposited by spin coating inside a nitrogen glovebox. A volume of 30 $\mu\mathrm{L}$ precursor solution was dispensed onto the substrate, followed by a 3 s acceleration step and spinning at 3000 rpm for 45 s. Chlorobenzene (350 $\mu\mathrm{L}$) was dropped at 20 s during the spin-coating process as an anti-solvent. The resulting MAPbBr$_3$ thin films had a thickness of approximately 110 nm.

\subsubsection*{Fabrication of nanoimprint master molds.}
Gold-on-glass nanoimprint master molds were fabricated using electron-beam lithography (EBL). Fused silica substrates (15 × 15 mm$^2$) were cleaned in piranha solution (H$_2$SO$_4$:H$_2$O$_2$ = 3:1), rinsed, and dried under nitrogen flow. A bilayer resist consisting of MMA (EL6, 120 nm at 6000 rpm) and PMMA 950 A2 (80 nm at 3000 rpm) was spin-coated, followed by baking at 180 $^\circ$C for 5 min.

To mitigate charging during EBL on insulating substrates, a 10 nm gold charge-dissipation layer was deposited. Patterns were defined using an EBPG5000+ system with an exposure dose of 250 $\mu\mathrm{C}\,\mathrm{cm}^{-2}$. The resist was developed in MIBK:IPA (1:3) for 1 min, followed by rinsing in IPA. A 2 nm Cr adhesion layer and 50 nm Au were deposited by electron-beam evaporation (LAB600). Lift-off was performed in remover 1165 overnight, followed by brief sonication in remover, acetone, and isopropanol.

\subsubsection*{Nanoimprint fabrication of metasurfaces.}
Nanoimprint parameters were optimized empirically for perovskite films. A pressure of 40 MPa was applied for 10 min on 15$\times$15 mm$^2$ samples using the master mold, followed by thermal annealing at 90 $^\circ$C for 10 min. This process yielded grating structures replicating the master mold geometry in inverse form. The fabrication scheme is provided in the Supplementary Information. The molds can be reused after cleaning with deionized water and drying.

\subsubsection*{Gas doping procedure.}
For gas-phase doping, a 18.5 cm diameter Petri dish was placed on a hot plate at 50 $^\circ$C. A volume of 40 $\mu\mathrm{L}$ of 36.3\% HCl was introduced into the dish, which was then sealed and allowed to equilibrate for 5 min to generate a saturated vapor environment. The perovskite metasurfaces were subsequently exposed to the HCl vapor for a controlled duration $t$. After exposure, the optical response of the samples was measured. Reverse doping was performed using HBr vapor under analogous conditions.

\subsection*{Angle-resolved spectroscopy}
Angle-resolved spectroscopy measurements were performed using a 4f optical system incorporating a back focal plane (BFP) imaging configuration. The setup consists of a slit spectrometer coupled to an EMCCD camera for detection. Broadband illumination was provided by a halogen lamp coupled into an optical fiber via a collimation lens. A Mitutoyo Plan Apo infinity-corrected objective (50$\times$, NA = 0.42) was used for both excitation and collection. Spatial filtering was implemented at the intermediate image plane within the 4\textit{f} system. Angle-resolved photoluminescence measurements were performed using the same optical configuration. Excitation was provided by a femtosecond laser with tunable wavelength (at 515 nm), 1 kHz repetition rate, and 200 fs pulse duration. The pump spot size (~15 $\mu$m) was defined by focusing the beam in the back focal plane using a lens with a 300 mm focal length, which does not affect the lasing spectra at working fluence lower than 140~$\mu$J/cm$^2$ for the undoped film (Figure~S9) caused by exciton-polaritons localization within the spot~\cite{masharin2026non}. The collected signal was directed to a CCD camera via a 100 mm tube lens placed after a beam splitter. This configuration enables imaging in both real and Fourier spaces.

\subsection*{Numerical modeling}

\subsubsection*{Spectra}
The eigenvalue problem and scattering problem were solved using the finite difference frequency domain (FDFD) method~\cite{rumpf2022electromagnetic}.
As shown in Fig.~\ref{fig1}e, periodic boundary conditions (PBCs) were applied at the transverse boundaries of the unit cell.
Dirichlet boundary conditions were applied at the top and bottom boundaries of the physical domain, while uniaxial perfectly matched layers (UPMLs) were employed to eliminate reflections from the Dirichlet boundaries.
A uniform grid with a cell size of $\Delta x = \Delta y = 2.5$ nm was used for modeling.
The total field/scattered field (TF/SF) technique was used to inject a TE-polarized plane wave source. 
The refractive indices of $\mathrm{SiO_2}$ and $\mathrm{Air}$ were assumed to be constant and equal to 1.5 and 1, respectively.
The refractive index of the perovskite after different doping times was measured by ellipsometry and used in the FDFD simulations, see SI.

\section*{Funding}

This research was supported by the Priority 2030 Federal Academic Leadership Program.

\bibliography{sample}

\noindent\textbf{Acknowledgments}\\
The authors acknowledge financial support from the National Natural Science Foundation of China (project 62350610272), the Department of Science and Technology of Shandong Province (Grant KY0020240040), and the Federal Academic Leadership Program Priority 2030. The work of PSP was financially supported by the ITMO Fellowship Program. The authors acknowledge the use of nanofabrication facilities at the Center of MicroNano Technology of \'Ecole Polytechnique F\'ed\'erale de Lausanne.\\

\noindent\textbf{Author contributions}\\
N.G. and M.M. fabricated the structures with input from H.A.. N.G. and Y.K. performed spectroscopy measurements. N.G., Y.K. developed the back focal plane spectroscopy setup. X.L. performed samples characterization. P.P, M.R.-B., D.M. and A.B. performed numerical simulations and developed the theoretical diffusion model. N.G., Y.K., P.P. and A.B. processed and analyzed the theoretical and experimental data. N.G., Y.K., P.P., M.M., M.R.-B., D.M., A.B. and S.M. wrote the manuscript with input from all authors. N.G., Y.K. and P.P designed the figures. A.B. and S.M. guided the research.\\

\noindent\textbf{Competing interests}\\
There are no competing interests to declare.

\noindent\textbf{Supplementary information (optional)}
If your article requires supplementary information, please include these files for peer-review. Please note that supplementary information will not be edited.

\end{document}